\documentclass[a4paper,11pt]{article}
\pdfoutput=1 

\usepackage{jheppub} 

\usepackage[T1]{fontenc} 

\usepackage{lscape}

\usepackage{multirow}

\usepackage[utf8]{inputenc}

\allowdisplaybreaks

\title{Classification of Robinson-Trautmann and Kundt Geometries with Large D Limit}

\author[a]{P\i nar Kirezli,\note{Corresponding author.}}

\affiliation[a]{Department of Physics, Faculty of Arts and Sciences, Nam\i k Kemal University,\\Tekirdag, 59030, Turkey}

\emailAdd{pkirezli@nku.edu.tr}

\abstract{Algebraic classification of higher dimensional, shear-free, twist-free, expanding (or non-expanding) spacetime is studied with the limit of $D\rightarrow\infty$. Similar to classification of any arbitrary dimension $D>4$, this spacetime is Type I(b) or more special, according to our calculations. However, thanks to the method of taking the limit of dimension $D\rightarrow \infty$, the components of Weyl scalar are obtained much simpler. Without solving field equations, by determining obligotary conditions to the components of Weyl scalar vanish, the spacetime is classified Type I(a), Type II(a-b-c-d), Type III(a-b), Type N and Type O for primary Weyl aligned null direciton (WAND), and Type $I_{i}$, Type $II_{i}$, Type $III_{i}$ and Type D(a-b-c-d) for secondary WAND.}

\arxivnumber{2201.13289}

\begin{document} 
\maketitle
\flushbottom

\section{Introduction}

Higher dimensional spacetimes are commonly studied to unify string theory and general relativity and to analyze the quantum field theories with CFT/AdS correspondence and to fully understand the efficiency of the theory of general relativity. Nearly during past decade, Emparan and \emph{et al}. examine the solutions of higher dimensional spacetimes in the limit of dimension $D\rightarrow\infty$ \cite{Emparan:2013moa,Emparan:2013xia,Emparan:2014cia,Emparan:2015gva,Emparan:2015rva,Andrade:2019edf,Licht:2020odx,Suzuki:2020kpx,Suzuki:2021lrw,Licht:2022wmz}. The obtained results make it easier to gain a new perspective on  theory of general relativity and especially analytical solutions, but it is also tested to be used in some other areas, i.e. quantum entanglement and holographics \cite{Colin-Ellerin:2019vst}. 

Robinson-Trautmann (RT) \cite{RT60,RT62} and Kundt \cite{Kundt61,Kundt62} spacetimes are special spacetimes which include many different solutions in 4-dimensions such as, pp-waves,  plane wave space-times, the Schwarzschild and Reissner–Nordstrom black holes, the C-metric, the Vaidya solution, photon rockets and their non-rotating generalisations. RT spacetime is described by the existence of an expanding, shearfree and twistfree congruence of null geodesics, while, Kundt geometry is defined as non-expanding case. However, higher-dimensional solutions are not as rich as 4-dimensional solutions because several important solutions can not generalize to higher dimensions. Whereas, extension of RT spacetime to higher dimensions were studied for any cosmological constant or aligned pure radiation \cite{Podolsky:2006du}, aligned electromagnetic fields \cite{Ortaggio:2007hs}, and general p-form fields \cite{Ortaggio:2014gma}. Additionally, the Kundt spacetime generalization to higher dimensions is obtained in \cite{Podolsky:2008ec}. 

Classification of spacetimes which enables to divide
the gravitational fields in an invariant way into distinct types, was first studied by Petrov \cite{Petrov}. There are several ways to obtain the classification of spacetimes such as,  null vectors, 2-spinors or scalar invariants. Since the Weyl scalar and cosmological constant include all data about curvature of a spacetime, the classification is fully understandable by studying components of Weyl scalar. Classification provided a better understanding of several aspects of General Relativity in 4-dimensions and it was extended to  any $D>4$ dimensions \cite{Milson:2004jx} and for RT \cite{Podolsky:2014mpa} and Kundt \cite{Podolsky:2013qwa} geometries. However, the algebraic classification of the RT and Kundt spacetimes with the limit of $D\rightarrow\infty$ have not been studied before.

Therefore, the main purpose of the paper is obtaining algebraic classifications of RT and Kundt spacetimes in the limit of $D\rightarrow \infty$ and compare the results with previous calculations of any dimension $D>4$ .

The paper is organized as; Section \ref{s2} includes a brief summary of shearless and twistless, expanding or non-expanding spacetimes. Primary and secondary Weyl-aligned null direction (WAND) are reviewed and the classification for components of boost weight are shown. Components of Weyl scalars are obtained with the limit of $D\rightarrow \infty$ and the different types of spacetimes are discussed. In Section \ref{s3} and \ref{s4} algebraic classification of the non-expanding Kundt spacetime and expanding RT spacetime are stated and several special cases are investigated for the limit of $D\rightarrow\infty$, respectively. Christoffel symbols, Riemann and Ricci tensor and Ricci scalar are calculated for the general expanding, shearless, twistfree spacetime for any dimension $D>4$ while the components of Weyl scalar are computed for the limit $D\rightarrow\infty$ in Appendix \ref{appendix}.

	\section{Non-Twisting, Shear-free Geometry }\label{s2}
In general, D-dimensional, shear-free and twist-free metric can be written in the form;
\begin{eqnarray}\label{metric}
	ds^2=g_{pq}\left(u,r,x\right)dx^{p}dx^q+2g_{up}\left(u,r,x\right)dudx^{p}-2du dr+g_{uu}\left(u,r,x\right)du^2 
\end{eqnarray}
where latin indices {p,q,...} count to $2$ to $(D-2)$ and $x$ is shorthand of these $D-2$ spatial coordinates on the traverse space. We can write some of the relations between contravariant and covariant metric components as;
\begin{eqnarray}
	g^{ur}=-1,~~~~g^{rp}=g^{pq}g_{uq},~~~~g^{rr}=-g_{uu}+g^{pq}g_{up}g_{uq},~~~~g_{up}=g^{rq}g_{pq}.\nonumber
\end{eqnarray}
Non-twisting spacetimes require a null hypersurface like as $u=$constant which is similar to existence of a null vector field \textbf{k} that is tangent to $u$=cons. surface in everywhere ( it is more common to choose \textbf{k} at $r$ direction and it becomes $\textbf{k}=\partial_r$.). This null vector covariant derivative for the metric (\ref{metric}) is obtained $k_{i;j}=\Gamma^{u}_{~i j}=\frac{1}{2}g_{i j,r}$ which satisfies $k_{r;j}=k_{i;r}=0$. Another orthonormal basis in the $D-2$ dimensional spatial coordinates can be defined as $m_i^{p}$ which is useful to identify optical matrix; $\rho_{i j}=k_{p;q}m_{i}^{p}m_{j}^{q}$. For shearless and twist-free spacetimes it reduces to;
$\rho_{i j}=\Theta \delta_{i j}$.
The expansion scalar is obtained $2 \Theta g_{pq}=g_{pq,r}$ by the definition of $\delta_{ij}=g_{pq}m^{p}_{i}m^{q}_{j}$ \cite{Podolsky:2014mpa}. The vanishing expansion, $\Theta=0$ is known the Kundt class \cite{Kundt61,Kundt62,Stephani,Griffits} which the spatial metric is independent of the affine parameter $r$ (in metric (\ref{metric}) $g_{pq}(u,r,x)\rightarrow g_{pq}(u,x)$).  Otherwise, expanding case, $\Theta\neq 0$ is named Robinson-Trautman class \cite{Stephani,Griffits,RT60,RT62}.

We begin with identify natural null frames to obtain classification of the spacetime which is given metric (\ref{metric}) as;
\begin{eqnarray}
	\mathbf{k}=\partial_{r} ~~~~\mathbf{\ell}=\frac{1}{2}g_{uu}\partial_{r}+\partial_{u}~~~~~\mathbf{m_i}=m_{i}^{p}\left(g_{up}\partial_{r}+\partial_{p}\right)
\end{eqnarray}
which satify the normalization conditions; $\mathbf{k}.\mathbf{\ell}=-1$, $\mathbf{m_i}.\mathbf{m_j}=\delta_{ij}$. Boosts defined as the rescaling of these null basis like; $k\rightarrow \lambda k$, $\ell\rightarrow \lambda^{-1} \ell $ and $m_{i}\rightarrow m_{i}$ and boost weight which is used to determine the classification of the spacetimes in higher dimensions, are obtained $+1,-1,0$, respectively. Additionally, if the boost weight $+2$ components of the Weyl tensor are zero, the null direction of $\mathbf{k}$  becomes a primary Weyl-aligned null direction (WAND) which is  analogue of a PND in 4-dimensions. Also, it will be multiple WAND if the $+1$ boost weight components of the Weyl scalar vanish and if the spacetime obeys multiple WAND conditions, it becomes algebraically special. One prepared the classification of the spacetime with obligatory vanishing components of boost weight \cite{Durkee:2011tx} which is summarized in Table \ref{table3}. Further, for fixed $\mathbf{k}$, $\mathbf{\ell}$ can be introduced a secondary WAND in order to obtain as many as null frame scalars vanish. According to secondary WAND the spacetime will be Type $I_{i}, II_{i}, III_{i}$ and Type D.
\begin{table}[h!]
	\centering
	\begin{tabular}{c|ccccc|c}
		\cline{2-6}
		& \multicolumn{5}{c|}{Boost Weight}                                                                            &                       \\ 
		&
		\multicolumn{5}{c|}{ Components}                                              &                       \\ \hline
		\multicolumn{1}{|c|}{Types} & \multicolumn{1}{c|}{+2} & \multicolumn{1}{c|}{+1} & \multicolumn{1}{c|}{0} & \multicolumn{1}{c|}{-1} & -2 & \multicolumn{1}{c|}{primary WAND} \\ \hline
		\multicolumn{1}{|c|}{G} & \multicolumn{1}{c|}{} & \multicolumn{1}{c|}{} & \multicolumn{1}{c|}{} & \multicolumn{1}{c|}{} &  & \multicolumn{1}{c|}{No} \\ \hline
		\multicolumn{1}{|c|}{I} & \multicolumn{1}{c|}{0} & \multicolumn{1}{c|}{} & \multicolumn{1}{c|}{} & \multicolumn{1}{c|}{} &  & \multicolumn{1}{c|}{a WAND} \\ \hline
		\multicolumn{1}{|c|}{II} & \multicolumn{1}{c|}{0} & \multicolumn{1}{c|}{0} & \multicolumn{1}{c|}{} & \multicolumn{1}{c|}{} &  & \multicolumn{1}{c|}{multiple WAND} \\ \hline
		\multicolumn{1}{|c|}{III} & \multicolumn{1}{c|}{0} & \multicolumn{1}{c|}{0} & \multicolumn{1}{c|}{0} & \multicolumn{1}{c|}{} &  & \multicolumn{1}{c|}{multiple WAND} \\ \hline
		\multicolumn{1}{|c|}{N} & \multicolumn{1}{c|}{0} & \multicolumn{1}{c|}{0} & \multicolumn{1}{c|}{0} & \multicolumn{1}{c|}{0} &  & \multicolumn{1}{c|}{multiple WAND} \\ \hline
		\multicolumn{1}{|c|}{O} & \multicolumn{1}{c|}{0} & \multicolumn{1}{c|}{0} & \multicolumn{1}{c|}{0} & \multicolumn{1}{c|}{0} & 0 & \multicolumn{1}{c|}{multiple WAND} \\ \hline
	\end{tabular}
	\caption{Higher dimensional classification of a spacetime with obligatory vanishing components of boost weights.}\label{table3}
\end{table}

Subsequently, as the dimension of the spacetime goes to infinity components of the Weyl scalar of the metric (\ref{metric}) become (Christoffel symbols, Riemann and Ricci tensors, Ricci scalar and Weyl tensor are calculated at the Appendix \ref{appendix}), 
\begin{eqnarray}
	\Psi_{0^{ij}}&=&C_{abcd}k^{a}m^{b}_ik^cm_j^d=m_{i}^{p}m_{j}^qC_{rprq}=0,\\
	\Psi_{1T^{i}}&=&C_{abcd}k^{a}\ell^bk^cm_i^d=m_i^pC_{rurp}=m_{i}^p\left[\left(-\frac{1}{2}g_{up,r}+\Theta g_{up}\right)_{,r}+\Theta_{,p}\right]~~\\
	\Psi_{1^{ijk}}&=&C_{abcd}k^{a}m_i^bm_j^cm_k^d=m_,^pm_j^qm_k^m C_{prmq}=0\\
	\Psi_{2S}&=&C_{abcd}k^{a}\ell^b\ell^ck^d=C_{ruur}=\frac{1}{4}g_{uu,rr}-\frac{1}{4}g^{pq}g_{up,r}g_{uq,r}-g_{uu}\Theta^2-g_{uu}\Theta_{,r}\nonumber\\
	&&+\Theta g_{uu,r}-\Theta g^{rp}g_{up,r}+2\Theta_{,u}\\
	\Psi_{2T^{(ij)}}&=&C_{abcd}k^{a}m_i^b\ell^cm_j^d=m_i^pm_j^q\left(C_{rpuq}+g_{up}C_{rpur}+\frac{1}{2}g_{uu}C_{rprq}\right)\nonumber\\
	&=&m_i^pm_j^q\bigg(\frac{1}{2}g_{pn}g^{ms}g_{us,r}~^s\Gamma^{n}_{~mq}+\frac{1}{4}g_{up,r}g_{uq,r}+\frac{1}{2}g_{pn}\left(g^{nm}g_{um,r}\right)_{,q}\nonumber\\
	&&-\frac{\Theta}{2}\Big(-g_{pq}g^{rm}g_{um,r}+g_{pq}g_{uu,r}+2g_{u(p}g_{q)u,r}+2E_{qp}\Big)\nonumber\\
	&&-g_{pq}\Theta_{,u}-\frac{g_{up}}{2}\Theta_{,q}+\Theta^2g_{up} g_{uq}-\frac{g^{rr}}{2}g_{pq,r}+\frac{g_{up}}{2}g_{up,rr}\nonumber\\
	&&-\Theta g_{up}g_{up,r}-\frac{g_{up}}{2}\Theta_{,p}-g_{up}^{~~2}\Theta_{,r}\bigg)\\
	\Psi_{2^{ijkl}}&=&C_{abcd}m_i^{a}m_j^bm_k^cm_l^d=m_i^pm_j^qm_k^nm_l^m\big(C_{pqmn}+g_{up}C_{rqmn}+g_{uq}C_{prmn}+g_{um}C_{pqrn}\nonumber\\
	&&+g_{un}C_{pqmr}\big)=m_i^pm_j^qm_k^nm_l^mC_{pqnm},\\
	\Psi_{2^{ij}}&=&C_{abcd}k^{a}\ell^b m_i^cm_j^d=m_i^p m_j^q\left(C_{rupq}+g_{uq}C_{rurp}+g_{up}C_{ruqr}\right)=m_i^p m_j^q \bigg(g_{u[p,q]r}\nonumber\\
	&&-2g_{u[p}\Theta_{,q]}+g_{u[p}g_{q]u,rr}+\Theta\left(2g_{u[q}g_{p]u,r}+E_{qm}-E_{pm}\right)\bigg),\\
	\Psi_{3T^{i}}&=&C_{abcd}\ell^{a}k^b\ell^cm_i^d=m_i^p \left(\frac{1}{2}g_{uu}C_{urrp}+g_{up}C_{urur}+C_{urup}\right)\nonumber\\
	&=&m_i^p\bigg(\frac{1}{2}g_{u[u}g_{p]u,rr}-g_{u[u,p]r}-2g_{up}\Theta_{,u}+\frac{g_{uu}g_{up}}{2}\Theta_{,r}+\frac{3g_{uu}}{4}\Theta_{,p}+\frac{1}{4}g^{rp}g_{up,r}g_{uq,r}\nonumber\\
	&&+\frac{\Theta}{2}\left(-g_{up}g_{uu,r}+2g_{up}g^{rp}g_{up,r}+g_{uu,p}+g^{rr}g_{up,r}+2g^{rs}E_{sp}\right)\nonumber\\
	&&+\frac{g_{um,r}}{4}\left(2g^{mn}E_{np}-g_{up,r}\right)\bigg)\\
	\Psi_{3^{ijk}}&=&C_{abcd}\ell^{a}m_i^bm_j^cm_k^d=m_i^pm_j^qm_k^m\bigg(\frac{1}{2}g_{uu}\left(C_{rpqm}+g_{uq}C_{rprm}+g_{um}C_{rpqr}\right)\nonumber\\
	&&+g_{up}\left(C_{urqm}+g_{uq}C_{urrm}+g_{um}C_{urqr}\right)+g_{uq}C_{uprm}+g_{um}C_{upqr}+C_{upqm}\bigg)\nonumber\\
	&=&m_i^pm_j^qm_k^m\bigg(\frac{1}{2}g_{uq}g_{mn}\left(g^{n\ell}g_{u\ell,r}\right)_{,q}-\frac{1}{2}g_{um}g_{qn}\left(g^{n\ell}g_{u\ell,r}\right)_{,p}+g_{up}g_{u[q}g_{m]u,rr}\nonumber\\
	&&-\frac{1}{2}g_{up,r}g_{u[q}g_{m]u,r}+\Theta_{,u}g_{u[q}g_{m]p}-g_{pn}\left(g^{rn}g_{u[q,r}\right)_{,m]}-\Theta\left(g^{r\ell}g_{u\ell,r}g_{u[q}g_{m]p}+g_{uu,r}g_{u[q}g_{m]p}\right)\nonumber\\
	&&-\Theta\left(g_{up}\left(E_{mn}-E_{qn}\right)+2E_{p[q}g_{m]u}\right)+E_{p[m}g_{q]u,r}-2g_{pn}\left(g^{n\ell}E_{\ell[q}\right)_{,m]}\nonumber\\
	&&+g_{pn}g^{s\ell}E_{\ell[q}~^s\Gamma^n_{~m]s}-g_{pn}g^{rs}~^s\Gamma^{n}_{~s[m}g_{q]u,r}+g^{\ell s}g_{us,r}~^s\Gamma^n_{~\ell p}g_{u[q}g_{m]n}\bigg)\\
	\Psi_{4^{ij}}&=&C_{abcd}\ell^{a}m_i^b\ell^cm_j^d=m_i^pm_j^q\bigg(\frac{g_{uu}}{2}\left(C_{rpuq}+C_{uprq+}\frac{g_{uu}}{2}C_{rprq}+g_{uq}C_{rpru}+g_{up}C_{urrq}\right)\nonumber\\
	&&+g_{up}\left(C_{uruq}+g_{uq}C_{urur}\right)+g_{uq}C_{upur}+C_{upuq}\bigg)=m_i^pm_j^q\bigg(\frac{1}{2}g_{up}g_{uq}g^{pq}g_{up,r}g_{uq,r}\nonumber\\
	&&+\frac{g_{uu}}{2}\bigg[g^{ms}g_{us,r}~^s\Gamma^{n}_{~m(p}g_{q)n}+\left(g^{mn}g_{um,r}\right)_{(q}g_{p)n}-2g_{pq} \Theta_{,u}-\Theta_{,(p}g_{q)u}+\Theta_{,[p}g_{q]u}\nonumber\\
	&&-g_{u[q}g_{p]u,rr}-\Theta\left(-g_{pq}g^{rm}g_{um,r}+g_{pq}g_{uu,r}+2g_{u(p}g_{q)u,r}+E_{qp}+E_{pq}\right)\bigg]\nonumber\\
	&&-g_{uu,r(p}g_{q)u}+g_{u(p}g_{q)u,r}+g_{up}g_{uq}g_{uu}\Theta_{,r}-\frac{1}{2}g_{um,r}g_{u(p}g_{q)u,r}-\frac{1}{4}g_{up}g_{uq}g_{uu,r}\nonumber\\
	&&+2g_{uu}\Theta_{,(p}g_{q)u}+g_{up}g_{u[u,q]r}+\frac{1}{4}g_{up,r}g_{uu,q}-2g_{up}g_{uq}\Theta_{,u}+g^{mn}g_{um,r}E_{n(q}g_{p)u}\nonumber\\
	&&+\Theta\left[-g_{up}E_{up}+g_{uq}g_{up,r}\left(g_{uu}+g_{up}g^{rp}\right)\right]+E_{p[q}g_{u]u,r}+\frac{1}{2}g^{rs}g_{uq,r}E_{ps}-g^{\ell s}E_{ps}E_{\ell q}\nonumber\\
	&&-\frac{1}{2}g^{r\ell}g_{up,r}E_{\ell q}-\frac{g_{pn}}{2}\left[\left(g^{rn}g_{uu,r}\right)_{,q}-\left(g^{rn}g_{uq,r}\right)_{,u}-2\left(g^{ns}E_{us}\right)_{,q}+2\left(g^{ns}E_{sq}\right)_{,u}\right]\bigg)~~
\end{eqnarray}
where $C_{pqnm}, E_{pq}, E_{up}$ equations are given in Appendix A. Analyzing of the classification for any dimension $D>4$ are studied in \cite{Podolsky:2014mpa}. 

As the Weyl scalar $\Psi_{0^{ij}}=0$, $\mathbf{k}$ is a primary WAND of the metric (\ref{metric}) and the spacetime is Type I. If $\Psi_{1T^{i}}$ vanishes, $\mathbf{k}$ becomes multiple WAND due to boost weight component of $+2, +1$ become zero. Meanwhile, vanishing components of the Weyl scalars determine the spacetime will become Type I(a)-I(b), Type II(a)-II(b)-II(c)-II(d) Type N and Type O and their combinations such as TypeII(ac) or Type II(bcd), for primary WAND $\textbf{k}$. For instant, the spacetime is called Type II(c) when the Weyl scalars $\Psi_{0^{ij}}=\Psi_{1T^{i}}=\Psi_{1^{ijk}}=\Psi_{2^{ijkl}}=0$ or Type III(a) when the scalars $\Psi_{0^{ij}}=\Psi_{1T^{i}}=\Psi_{1^{ijk}}=\Psi_{2S}=\Psi_{2T^{ij}}=\Psi_{2^{ijkl}}=\Psi_{2^{ij}}=\Psi_{3T^{i}}=0$. Because of the components of the Weyl scalar $\Psi_{o^{ij}}=\Psi_{1^{ijk}}=0$, the spacetime of shear-free, twist-free, expanding (or not expanding) can be classified Type I(b) or more special. Withal, metric (\ref{metric}) spacetime is not algebraically special, because of all $+1$ components of boost weight are not zero ($\Psi_{1T^{i}}\neq 0$).

Furthermore, the secondary WAND is the natural null vector $\mathbf{\ell}$ which classifies the spacetime as the Type $I_i$~-$II_i$~-$III_{i}$ and Type D, such as, the spacetime will be Type $II_i$, when $\Psi_{4^{ij}}=0$ addition to $\Psi_{0^{ij}}=\Psi_{1T^{i}}=\Psi_{1^{ijk}}=0$ or the spacetime becomes Type D(a) when the Weyl tensors $\Psi_{0^{ij}}=\Psi_{1T^{i}}=\Psi_{1^{ijk}}=\Psi_{2S}=\Psi_{3T^{i}}=\Psi_{3^{ijk}}=\Psi_{4^{ij}}=0$. One prepared Table \ref{table1} for the relationship between vanishing Weyl scalars and the spacetimes' classification, with both WANDs $\mathbf{k}$ and $\mathbf{\ell}$ \cite{Podolsky:2014mpa}. 

While the number of  dimension goes to infinity, the classification of metric (\ref{metric}) did not change, just the equations became simpler than the functions were obtained for any dimensions $D>4$. It is the power of the large $D$ expansion method which allows to obtain analytical solutions for classification of this spacetime. Hereafter, we will determine (sub)-types of the higher dimensional, shear-free, twist-free and expanding or non-expanding spacetime for special conditions which will correspond specific spacetimes. 

\section{Kundt Spacetime}\label{s3}
Kundt spacetimes are shear-free, twist-free, non-expanding geometries which will be obtained by taking $\Theta=0$ and the metric functions of spatial coordinates independent of the parameter $r$ at the metric (\ref{metric}). While the algebraic classification of the Kundt geometry for higher dimensions has been studied in \cite{Podolsky:2008ec,Podolsky:2014mpa,Podolsky:2013qwa}, we will classified this geometry as the number of dimensions goes to infinity.
Kundt spacetimes will be Type I(b) ($\Psi_{0^{ij}}=\Psi_{1^{ijk}}=0$) or more special which is determined by the vanishing Weyl scalars. If we set the spacetime Type I(a) (it becomes Type II too), the Weyl scalar $\Psi_{1T^{i}}$ have to vanish. As a result of it, the spacetime becomes algebraically special because all +1 component of the boost weight vanish and the metric function $g_{up,rr}=0$ and it reads;

\begin{landscape}
	\begin{table}[]
		\begin{tabular}{|c|ccccccccccccccccc|}
			\hline
			\multirow{3}{*}{Components of Weyl Scalar} & \multicolumn{17}{c|}{ Types}                                                                                                                                                                                                                                                                                                                                                                           \\ \cline{2-18} 
			& \multicolumn{13}{c|}{WAND $\mathbf{k}$}                                                                                                                                                                                                                                                                                                & \multicolumn{4}{c|}{WAND $\mathbf{\ell}$}                                                    \\ \cline{2-18} 
			& \multicolumn{1}{c|}{I} & \multicolumn{1}{c|}{I(a)} & \multicolumn{1}{c|}{I(b)} & \multicolumn{1}{c|}{II} & \multicolumn{1}{c|}{II(a)} & \multicolumn{1}{c|}{II(b)} & \multicolumn{1}{c|}{II(c)} & \multicolumn{1}{c|}{II(d)} & \multicolumn{1}{c|}{III} & \multicolumn{1}{c|}{III(a)} & \multicolumn{1}{c|}{III(b)} & \multicolumn{1}{c|}{N} & \multicolumn{1}{c|}{O} & \multicolumn{1}{c|}{$I_{i}$} & \multicolumn{1}{c|}{$II_{i}$} & \multicolumn{1}{c|}{$III_{i}$} & D \\ \hline
			$\Psi_{0^{ij}}$	& \multicolumn{1}{c|}{0} & \multicolumn{1}{c|}{0} & \multicolumn{1}{c|}{0} & \multicolumn{1}{c|}{0} & \multicolumn{1}{c|}{0} & \multicolumn{1}{c|}{0} & \multicolumn{1}{c|}{0} & \multicolumn{1}{c|}{0} & \multicolumn{1}{c|}{0} & \multicolumn{1}{c|}{0} & \multicolumn{1}{c|}{0} & \multicolumn{1}{c|}{0} & \multicolumn{1}{c|}{0} & \multicolumn{1}{c|}{0} & \multicolumn{1}{c|}{0} & \multicolumn{1}{c|}{0} & 0 \\ \hline
			$\Psi_{1T^{i}}$	& \multicolumn{1}{c|}{} & \multicolumn{1}{c|}{0} & \multicolumn{1}{c|}{} & \multicolumn{1}{c|}{0} & \multicolumn{1}{c|}{0} & \multicolumn{1}{c|}{0} & \multicolumn{1}{c|}{0} & \multicolumn{1}{c|}{0} & \multicolumn{1}{c|}{0} & \multicolumn{1}{c|}{0} & \multicolumn{1}{c|}{0} & \multicolumn{1}{c|}{0} & \multicolumn{1}{c|}{0} & \multicolumn{1}{c|}{} & \multicolumn{1}{c|}{0} & \multicolumn{1}{c|}{0} & 0 \\ \hline
			$\Psi_{1^{ijk}}$ & \multicolumn{1}{c|}{} & \multicolumn{1}{c|}{} & \multicolumn{1}{c|}{0} & \multicolumn{1}{c|}{0} & \multicolumn{1}{c|}{0} & \multicolumn{1}{c|}{0} & \multicolumn{1}{c|}{0} & \multicolumn{1}{c|}{0} & \multicolumn{1}{c|}{0} & \multicolumn{1}{c|}{0} & \multicolumn{1}{c|}{0} & \multicolumn{1}{c|}{0} & \multicolumn{1}{c|}{0} & \multicolumn{1}{c|}{} & \multicolumn{1}{c|}{0} & \multicolumn{1}{c|}{0} & 0 \\ \hline
			$\Psi_{2S}$		& \multicolumn{1}{c|}{} & \multicolumn{1}{c|}{} & \multicolumn{1}{c|}{} & \multicolumn{1}{c|}{} & \multicolumn{1}{c|}{0} & \multicolumn{1}{c|}{} & \multicolumn{1}{c|}{} & \multicolumn{1}{c|}{} & \multicolumn{1}{c|}{0} & \multicolumn{1}{c|}{0} & \multicolumn{1}{c|}{0} & \multicolumn{1}{c|}{0} & \multicolumn{1}{c|}{0} & \multicolumn{1}{c|}{} & \multicolumn{1}{c|}{} & \multicolumn{1}{c|}{0} &  \\ \hline
			$\Psi_{2T^{ij}}$ & \multicolumn{1}{c|}{} & \multicolumn{1}{c|}{} & \multicolumn{1}{c|}{} & \multicolumn{1}{c|}{} & \multicolumn{1}{c|}{} & \multicolumn{1}{c|}{0} & \multicolumn{1}{c|}{} & \multicolumn{1}{c|}{} & \multicolumn{1}{c|}{0} & \multicolumn{1}{c|}{0} & \multicolumn{1}{c|}{0} & \multicolumn{1}{c|}{0} & \multicolumn{1}{c|}{0} & \multicolumn{1}{c|}{} & \multicolumn{1}{c|}{} & \multicolumn{1}{c|}{0} &  \\ \hline
			$\Psi_{2^{ijkl}}$	& \multicolumn{1}{c|}{} & \multicolumn{1}{c|}{} & \multicolumn{1}{c|}{} & \multicolumn{1}{c|}{} & \multicolumn{1}{c|}{} & \multicolumn{1}{c|}{} & \multicolumn{1}{c|}{0} & \multicolumn{1}{c|}{} & \multicolumn{1}{c|}{0} & \multicolumn{1}{c|}{0} & \multicolumn{1}{c|}{0} & \multicolumn{1}{c|}{0} & \multicolumn{1}{c|}{0} & \multicolumn{1}{c|}{} & \multicolumn{1}{c|}{} & \multicolumn{1}{c|}{0} &  \\ \hline
			$\Psi_{2^{ij}}$	& \multicolumn{1}{c|}{} & \multicolumn{1}{c|}{} & \multicolumn{1}{c|}{} & \multicolumn{1}{c|}{} & \multicolumn{1}{c|}{} & \multicolumn{1}{c|}{} & \multicolumn{1}{c|}{} & \multicolumn{1}{c|}{0} & \multicolumn{1}{c|}{0} & \multicolumn{1}{c|}{0} & \multicolumn{1}{c|}{0} & \multicolumn{1}{c|}{0} & \multicolumn{1}{c|}{0} & \multicolumn{1}{c|}{} & \multicolumn{1}{c|}{} & \multicolumn{1}{c|}{0} &  \\ \hline
			$\Psi_{3T^{i}}$	& \multicolumn{1}{c|}{} & \multicolumn{1}{c|}{} & \multicolumn{1}{c|}{} & \multicolumn{1}{c|}{} & \multicolumn{1}{c|}{} & \multicolumn{1}{c|}{} & \multicolumn{1}{c|}{} & \multicolumn{1}{c|}{} & \multicolumn{1}{c|}{} & \multicolumn{1}{c|}{0} & \multicolumn{1}{c|}{} & \multicolumn{1}{c|}{0} & \multicolumn{1}{c|}{0} & \multicolumn{1}{c|}{} & \multicolumn{1}{c|}{} & \multicolumn{1}{c|}{} & 0 \\ \hline
			$\Psi_{3^{ijk}}$	& \multicolumn{1}{c|}{} & \multicolumn{1}{c|}{} & \multicolumn{1}{c|}{} & \multicolumn{1}{c|}{} & \multicolumn{1}{c|}{} & \multicolumn{1}{c|}{} & \multicolumn{1}{c|}{} & \multicolumn{1}{c|}{} & \multicolumn{1}{c|}{} & \multicolumn{1}{c|}{} & \multicolumn{1}{c|}{0} & \multicolumn{1}{c|}{0} & \multicolumn{1}{c|}{0} & \multicolumn{1}{c|}{} & \multicolumn{1}{c|}{} & \multicolumn{1}{c|}{} &  0\\ \hline
			$\Psi_{4^{ij}}$	& \multicolumn{1}{c|}{} & \multicolumn{1}{c|}{} & \multicolumn{1}{c|}{} & \multicolumn{1}{c|}{} & \multicolumn{1}{c|}{} & \multicolumn{1}{c|}{} & \multicolumn{1}{c|}{} & \multicolumn{1}{c|}{} & \multicolumn{1}{c|}{} & \multicolumn{1}{c|}{} & \multicolumn{1}{c|}{} & \multicolumn{1}{c|}{} & \multicolumn{1}{c|}{0} & \multicolumn{1}{c|}{0} & \multicolumn{1}{c|}{0} & \multicolumn{1}{c|}{0} & 0  \\ \hline
		\end{tabular}
			\caption{General algebraic classifications of shearfree, twistless spacetimes with necessary vanishing Weyl scalars for both WANDs $\mathbf{k}$ and $\mathbf{\ell}$ .}\label{table1}
	\end{table}
	
\end{landscape}

\begin{eqnarray*}
	g_{up}=rd_p(u,x)+c_p(u,x).
\end{eqnarray*}
If we keep going to determine the metric functions by the vanishing Weyl scalars for classification, we get that from $\Psi_{2S}=0$,
\begin{eqnarray*}
	g_{uu}=r^2g^{pq}(u,x)d_p(u,x)d_{q}(u,x)+rl(u,x)+s(u,x)
\end{eqnarray*}
where the solution is Type II(a) and the metric of Kundt spacetime becomes;
\begin{eqnarray}\label{metric2}
	ds^2&=&g_{pq}\left(u,x\right)dx^{p}dx^q+2\left(rd_p(u,x)+c_p(u,x)\right)dudx^{p}-2du dr\nonumber\\
	&+&\left(r^2g^{pq}(u,x)d_p(u,x)d_{q}(u,x)+rl(u,x)+s(u,x)\right)du^2.
\end{eqnarray}
Also, it easily becomes Type II(ad) by $\Psi_{2^{ij}}=0$, which gives $d_p(u,x)=d_p(u)$. However, the other subtypes are not simply obtained by only vanishing components of Weyl scalar. Therefore, we will analyze several cases which is obtained by simplification of metric components.
	\subsection{\textbf{All metric functions are independent of parameter $r$. }(Corresponding pp-waves.)} 
	
	pp-waves are one of the important subclass of the Kundt spacetimes. They are defined in Brinkmann \cite{Brinkmann:1925fr,Podolsky:2008ec,Podolsky:2013qwa} form by the metric (\ref{metric2}), as the all metric functions independent of the parameter $r$;
	\begin{eqnarray}
		ds^2&=&g_{pq}\left(u,x\right)dx^{p}dx^q+2c_p(u,x)dudx^{p}-2du dr+s(u,x)du^2.
	\end{eqnarray}
	With respect to independence of $r$, higher dimensional pp-waves can classified as Type II(abd) ($\Psi_{0^{ij}}=\Psi_{1T^{i}}=\Psi_{1^{ijk}}=\Psi_{2S}=\Psi_{2T^{(ij)}}=\Psi_{2^{ij}}=0$) or more special when the dimension number of the spacetime goes to infinity. Furthermore, if we set $~^sR_{pqnm}=c_{s}(u,x)~^s\Gamma^{s}_{~p[n}g_{m]q}$ ($\Psi_{2^{ijkl}}=0$) the pp-waves become Type III(a) and more special for multiple WAND $\mathbf{k}$. To make the pp-waves Type III(b) (simultaneously it becomes Type N), the condition of Weyl scalar $\Psi_{3T^{i}}=0$, gives that;
	\begin{eqnarray*}
		\left(g^{n\ell}E_{\ell[q}\right)_{,m]}=\frac{1}{2}g^{s\ell} E_{\ell [q}~^s\Gamma^{n}_{m]s}.
	\end{eqnarray*}
	There is one more condition which makes pp-waves Type O for WAND $\mathbf{k}$ is $\Psi_{4^{ij}}=0$ ;
	\begin{eqnarray*}
		g^{\ell s}E_{ps}E_{\ell q}+g_{pn}\left(-\left(g^{ns}E_{us}\right)_{,q}+\left(g^{ns}E_{sq}\right)_{,u}\right)=0.
	\end{eqnarray*}
	Only with this condition we can classified pp-waves Type $I_{i}$ and $II_{i}$ for secondary WAND $\mathbf{\ell}$. Higher dimensional pp-waves classification and obligatory conditions are summarized in Table \ref{table2}.
	
	\begin{table}[h]
		\begin{tabular}{|c|cc|c|}
			\hline
			Types for 	& \multicolumn{2}{c|}{Obligatory Conditions}    & 	Types for \\ 
			WAND $\mathbf{k}$	& \multicolumn{2}{c|}{}    &  WAND $\mathbf{\ell}$ \\\hline
			I	& \multicolumn{1}{c|}{always} & $E_{p[u,q]}=g^{\ell s}E_{ps}E_{\ell q}+g_{pn}g^{ns}_{~~~[,q}E_{s]u} $ & $I_i$ \\ \hline
			I(a)	& \multicolumn{1}{c|}{always} & $E_{p[u,q]}=g^{\ell s}E_{ps}E_{\ell q}+g_{pn}g^{ns}_{~~~[,q}E_{s]u} $ &$I(a)_i$  \\ \hline
			I(b) 	& \multicolumn{1}{c|}{always}	& $E_{p[u,q]}=g^{\ell s}E_{ps}E_{\ell q}+g_{pn}g^{ns}_{~~~[,q}E_{s]u} $ &  $I(b)_i$\\  \hline
			II	& \multicolumn{1}{c|}{always} &  $E_{p[u,q]}=g^{\ell s}E_{ps}E_{\ell q}+g_{pn}g^{ns}_{~~~[,q}E_{s]u} $ & $II_{i}$ \\  \hline
			II(a)	& \multicolumn{1}{c|}{always} &  $E_{p[u,q]}=g^{\ell s}E_{ps}E_{\ell q}+g_{pn}g^{ns}_{~~~[,q}E_{s]u} $ & $II(a)_{i}$ \\ \hline
			II(b)	& \multicolumn{1}{c|}{always} &  $E_{p[u,q]}=g^{\ell s}E_{ps}E_{\ell q}+g_{pn}g^{ns}_{~~~[,q}E_{s]u} $ & $II(b)_{i}$ \\ \hline
			II(c)	& \multicolumn{1}{c|}{$~^sR_{pqnm}=c_{s}(u,x)~^s\Gamma^{s}_{~p[n}g_{m]q}$} & $~^sR_{pqnm}=c_{s}(u,x)~^s\Gamma^{s}_{~p[n}g_{m]q}$ &  $II(c)_{i}$ \\
			&\multicolumn{1}{c|}{} &  $E_{p[u,q]}=g^{\ell s}E_{ps}E_{\ell q}+g_{pn}g^{ns}_{~~~[,q}E_{s]u} $ &  \\ \hline
			II(d)	& \multicolumn{1}{c|}{always} &  $E_{p[u,q]}=g^{\ell s}E_{ps}E_{\ell q}+g_{pn}g^{ns}_{~~~[,q}E_{s]u} $  &  $II(d)_{i}$\\ \hline
			III	& \multicolumn{1}{c|}{$~^sR_{pqnm}=c_{s}(u,x)~^s\Gamma^{s}_{~p[n}g_{m]q}$} &$~^sR_{pqnm}=c_{s}(u,x)~^s\Gamma^{s}_{~p[n}g_{m]q}$  & $ III_{i} $ \\
			& \multicolumn{1}{c|}{} & $E_{p[u,q]}=g^{\ell s}E_{ps}E_{\ell q}+g_{pn}g^{ns}_{~~~[,q}E_{s]u} $ &  \\ \hline
			III(a)	& \multicolumn{1}{c|}{$~^sR_{pqnm}=c_{s}(u,x)~^s\Gamma^{s}_{~p[n}g_{m]q}$} & $~^sR_{pqnm}=c_{s}(u,x)~^s\Gamma^{s}_{~p[n}g_{m]q}$ & $III(a)_{i}$ \\ 
			& \multicolumn{1}{c|}{} &  $E_{p[u,q]}=g^{\ell s}E_{ps}E_{\ell q}+g_{pn}g^{ns}_{~~~[,q}E_{s]u} $ &  \\\hline
			III(b)	& \multicolumn{1}{c|}{$~^sR_{pqnm}=c_{s}(u,x)~^s\Gamma^{s}_{~p[n}g_{m]q,}$} & $~^sR_{pqnm}=c_{s}(u,x)~^s\Gamma^{s}_{~p[n}g_{m]q,}$ &  $III(b)_{i}$ \\
			& \multicolumn{1}{c|}{$\left(g^{n\ell}E_{\ell[q}\right)_{,m]}=\frac{1}{2}g^{s\ell} E_{\ell [q}~^s\Gamma^{n}_{m]s} $}& $\left(g^{n\ell}E_{\ell[q}\right)_{,m]}=\frac{1}{2}g^{s\ell} E_{\ell [q}~^s\Gamma^{n}_{m]s} $ & \\
			&\multicolumn{1}{c|}{} & $E_{p[u,q]}=g^{\ell s}E_{ps}E_{\ell q}+g_{pn}g^{ns}_{~~~[,q}E_{s]u} $ &  \\  \hline
			N	& \multicolumn{1}{c|}{$~^sR_{pqnm}=c_{s}(u,x)~^s\Gamma^{s}_{~p[n}g_{m]q,}$} &$\left(g^{n\ell}E_{\ell[q}\right)_{,m]}=\frac{1}{2}g^{s\ell} E_{\ell [q}~^s\Gamma^{n}_{m]s} $  &  D=\\
			& \multicolumn{1}{c|}{$\left(g^{n\ell}E_{\ell[q}\right)_{,m]}=\frac{1}{2}g^{s\ell} E_{\ell [q}~^s\Gamma^{n}_{m]s} $}&$E_{p[u,q]}=g^{\ell s}E_{ps}E_{\ell q}+g_{pn}g^{ns}_{~~~[,q}E_{s]u} $ & (D(abd)) \\  \hline
			& \multicolumn{1}{c|}{$~^sR_{pqnm}=c_{s}(u,x)~^s\Gamma^{s}_{~p[n}g_{m]q,}$} & $~^sR_{pqnm}=c_{s}(u,x)~^s\Gamma^{s}_{~p[n}g_{m]q,}$  &  \\
			O	& \multicolumn{1}{c|}{$\left(g^{n\ell}E_{\ell[q}\right)_{,m]}=\frac{1}{2}g^{s\ell} E_{\ell [q}~^s\Gamma^{n}_{m]s} $,}& $\left(g^{n\ell}E_{\ell[q}\right)_{,m]}=\frac{1}{2}g^{s\ell} E_{\ell [q}~^s\Gamma^{n}_{m]s} $,  & D(c) \\  
			& \multicolumn{1}{c|}{$E_{p[u,q]}=g^{\ell s}E_{ps}E_{\ell q}+g_{pn}g^{ns}_{~~~[,q}E_{s]u} $}& $E_{p[u,q]}=g^{\ell s}E_{ps}E_{\ell q}+g_{pn}g^{ns}_{~~~[,q}E_{s]u} $ &  \\  
			\hline
		\end{tabular}
		\caption{Higher dimensional pp-waves classification with obligatory conditions for both WANDs  $\textbf{k}$ and $\mathbf{\ell}$ as the dimension of the spacetime $D\rightarrow \infty$.} \label{table2}
	\end{table}
	
	\subsection{$\mathbf{g_{up}}=0, \mathbf{g_{uu}}=\mathbf{r l(u,x)}$} 
	
	According to this special case, the solution becomes Type II(abd) or more special because of $\Psi_{2T^{(ij)}}=\Psi_{2^{ij}}=0$. Meanwhile, the spacetime will be Type II(c) as the Weyl scalar $\Psi_{2^{ijkl}}=0$ which is provided by taking flat spatial spacetime $~^sR_{pqnm}=0$ . Interestingly, when the metric function is $g_{uu,pr}=0$ which is obtained by $\Psi_{3^{ijk}}=0$ the spacetime become Type III(a). This can summarized as;
	\begin{equation*}
		l(u,x)\rightarrow l(u).
	\end{equation*}
	Moreover, this spacetime becomes Type III(b) (and also Type N), and Type O with respect to multiple WAND $\mathbf{k}$ if the Weyl scalars become zero as,
	\begin{equation*}
		\Psi_{3^{ijk}}=0\rightarrow \left(g^{n\ell}g_{\ell q,u}\right)_{,m}-\frac{g^{s\ell}}{2}g_{\ell q,u}~^s\Gamma^{n}_{~ms}=\left(g^{n\ell}g_{\ell m,u}\right)_{,q}-\frac{g^{s\ell}}{2}g_{\ell m,u}~^s\Gamma^{n}_{~qs}
	\end{equation*}
	\begin{equation*}
		\Psi_{4^{ij}}=0\rightarrow g_{pn}\left(g^{ns}g_{sq,u}\right)_{u}=\frac{l(u)}{2}g_{pq,u}-\frac{g^{\ell s}}{2}g_{ps,u}g_{\ell q,u}.
	\end{equation*}

	Additionally, algebraic classification for secondary WAND $\mathbf{\ell}$ can be written Type $I_{i}$ and Type $II_{i}$ if the condition $\Psi_{4^{ij}}=0$ is satisfied which gives;
	\begin{equation*}
		\Psi_{4^{ij}}=0\rightarrow g_{pn}\left[\left(g^{ns}g_{sq,u}\right)_{,u}+\left(g^{ns}g_{uu,p}\right)_{,q}\right]=\frac{l(u,x)}{2}g_{pq,u}-\frac{g^{\ell s}}{2}g_{ps,u}g_{\ell q,u}.
	\end{equation*}
	With this condition, if $~^sR_{pqnm}=0$ which means spatial spacetime is flat and the classification becomes Type $III_{i}$ for WAND $\mathbf{\ell}$. On the other hand, if we want to get Type D solutions, we have to set 
	\begin{eqnarray*}
		&&\Psi_{3^{ijk}}=0\rightarrow l(u,x)\rightarrow l(u)\\
		&&\Psi_{3^{ijk}}=0\rightarrow \left(g^{n\ell}g_{\ell q,u}\right)_{,m}-\frac{g^{s\ell}}{2}g_{\ell q,u}~^s\Gamma^{n}_{~ms}=\left(g^{n\ell}g_{\ell m,u}\right)_{,q}-\frac{g^{s\ell}}{2}g_{\ell m,u}~^s\Gamma^{n}_{~qs}\\
		&&\Psi_{4^{ij}}=0\rightarrow g_{pn}\left(g^{ns}g_{sq,u}\right)_{u}=\frac{l(u)}{2}g_{pq,u}-\frac{g^{\ell s}}{2}g_{ps,u}g_{\ell q,u}.
	\end{eqnarray*}
	Also the spacetime becomes Type D(abd) with these conditions in this case.

\section{Robinson-Trautman Spacetime}\label{s4}
Higher dimensional Robinson-Trautman geometry is defining non-twist, shearfree spacetimes, which expanding along the direction $r$ in metric (\ref{metric}). General algebraic classification of RT in higher dimensions as dimension numbers go to infinity is given in Table \ref{table1}. In here we will discuss some special cases. First, Riemannian Type I and Ricci Type I will be studied which is corresponding to expansion scalar $\Theta=\frac{1}{r}$ and $R_{rprq}=R_{rr}=0$. The spatial metric of the spacetime becomes;
\begin{equation*}
	g_{pq}=r^2h_{pq}(u,x).
\end{equation*}
According to this expansion scalar the spacetime is Type I and Type I(b) too, as $D\rightarrow \infty$. This spacetime becomes Type I(a) and also Type II and more special if the $\Psi_{1T^{i}}=0$ which is obtained by;
\begin{equation*}
	g_{up}=r^2d_{p}(u,x)+rc_{p}(u,x).
\end{equation*}
By using these metric functions, we can rewrite the components of Weyl scalar;
\begin{eqnarray}
	\Psi_{2S}&=&\frac{1}{4}g_{uu,rr}+\frac{g_{uu,r}}{r}-\left(2rd_{p}+c_{p}\right)\left(\frac{g^{pq}}{4}\left(2rd_{q}+c_{q}\right)+\frac{g^{rp}}{r}\right)\\
	\Psi_{2T^{ij}}&=&m_i^pm_j^q\Bigg[\frac{r^2h_{pn}}{2}\left[g^{ms}~^s\Gamma^n_{~mq}\left(2 r d_s+c_{s}\right)+2\left(g^{nm}\left(2rd_m+c_m\right)\right)_{,q}\right]\nonumber\\
	&&-\frac{E_{qp}}{2}+\frac{1}{4}c_pc_q-rg^{rr}h_{pq}\Bigg]\\
	\Psi_{2^{ijk\ell}}&=&m_i^pm_j^qm_k^nm_l^m\tilde{C}_{pqnm}\\
	\Psi_{2^{ij}}&=&\frac{1}{2}m_i^pm_j^q\left[\left(2rd_{p}+c_p\right)_{,q}-\left(2rd_{q}+c_q\right)_{,p}+\frac{2}{r}\left(E_{qm}-E_{pm}\right)\right]\\
	\Psi_{3T^{i}}&=&m_i^p\Bigg[-\frac{1}{2}g_{uu}\left(d_p+\frac{2c_p}{r}\right)-\frac{1}{4}g_{uu,rr}\left(r^2d_{p}+rc_p\right)-\frac{1}{2}g_{uu,pr}+\frac{g_{uu,r}}{2}\left(rd_p+c_p\right)\nonumber\\
	&&+\frac{g_{uu,p}}{2r}+\frac{1}{2}\left(2rd_p+c_p\right)_{,u}+g^{rr}\left(d_p+\frac{c_p}{2r}\right)+\frac{g^{rs}}{r}E_{sp}+\frac{g^{mn}}{2}\left(2rd_m+c_m\right)E_{np}\nonumber\\
	&&+g^{rp}\left(r^2d_pd_q+\frac{rd_pc_q}{2}+\frac{rd_qc_p}{2}+\frac{c_pc_q}{4}+2rd_p^2+3rd_pc_p+c_p^2\right)\nonumber\\
	&&-\left(r^2d_md_p+\frac{rd_mc_p}{2}+\frac{rd_pc_m}{2}-\frac{c_mc_p}{4}\right)\Bigg]\\
	\Psi_{3^{ijk}}&=&m_i^pm_j^qm_k^n\Bigg[\frac{r^3}{2}\left(h_{mn}\left(rd_q+c_q\right)\left(g^{n\ell}\left(2rd_{\ell}+c_{\ell}\right)\right)_{,q}-h_{qn}\left(rd_m+c_m\right)\left(g^{n\ell}\left(2rd_{\ell}+c_{\ell}\right)\right)_{,p}\right)\nonumber\\
	&&-r^2\left(rd_p+\frac{3}{2}c_p\right)d_{[q}c_{m]}-r^2\left(rd_{[q}h_{m]p}+c_{[q}h_{m]p}\right)\left(g^{r\ell}\left(2rd_{\ell}+c_{\ell}\right)+g_{uu,r}\right)\nonumber\\
	&&-r^2h_{pn}\left(\left(2rg^{rn}d_{[q}\right)_{,m]}+\left(g^{rn}c_{[q}\right)_{,m]}\right)-\left(rd_p+c_p\right)\left(E_{mn}-E_{qn}\right)-4rE_{p[q}d_{m]}-3E_{p[m}c_{q]}\nonumber\\
	&&+r^2h_{pn}\left(-2\left(g^{n\ell}E_{\ell[q}\right)_{,m]}+g^{s\ell}E_{\ell[q}~^s\Gamma^n_{~m]s}-g^{rs}~^s\Gamma^{n}_{~s[m}g_{q]u,r}\right)\nonumber\\
	&&+\frac{r^3g^{\ell s}}{2}\left(2rd_s+c_s\right)~^s\Gamma^n_{~\ell p}\left(rd_{[q}h_{m]n}+c_{[q}h_{m]n})\right)\bigg]\\
	\Psi_{4^{ij}}&=&  \Bigg[\frac{r^2g^{pq}}{2}\left(2r^2d_p^2+3rd_pc_p+c_p^2\right)\left(2r^2d_q^2+3rd_qc_q+c_q^2\right)+\frac{g_{uu}}{2}\big[r^2g^{ms}\left(2rd_s+c_s\right)~^s\Gamma^n_{~m(p}h_{q)n}\nonumber\\
	&&+r^2\left(g^{mn}\left(2rd_m+c_m\right)\right)_{(,q}h_{p)n}-2\left(r^2d_pd_q+rd_pc_q+2rd_qc_p+c_pc_q\right)-\frac{1}{r}\left(E_{qp}+E_{pq}\right)\nonumber\\
	&&-rh_{pq}\left(g_{uu,r}-rg^{rm}\left(rd_m+c_m\right)\right)\big]-\frac{r^2d_p+rc_p}{2}\left(\frac{rg_{uu,r}}{2}\left(rd_q+c_q\right)+\left(2rd_q+c_q\right)_{,u}\right)\nonumber\\
	&&+\left(\frac{r}{2}-\frac{r^2}{4}\left(rd_m+c_m\right)\right)\left(4r^2d_pd_q+3rd_pc_q+3rd_qc_p+2c_pc_q\right)\nonumber\\
	&&+\left(2rd_p+c_p\right)\left(\frac{g_{uu,q}}{4}+rg^{rp}\left(rd_p+c_p\right)\left(rd_q+c_q\right)-\frac{g^{r\ell}E_{\ell q}}{2}\right)\nonumber\\
	&&+\frac{rg^{mn}}{2}\left(2rd_m+c_m\right)\left(\left(rd_p+c_p\right)E_{nq}+\left(rd_q+c_q\right)E_{np}\right)\nonumber\\
	&&-\frac{rg_{uu,rp}\left(rd_q+c_q\right)}{2}+\frac{1}{2}E_{pq}g_{uu,r}-\frac{E_{up}}{2}\left(2rd_q+c_q\right)+\frac{1}{2}g^{rs}\left(2rd_q+c_q\right)E_{ps}-g^{\ell s}E_{ps}E_{\ell q}\nonumber\\
	&&-\frac{r^2h_{pn}}{2}\left[\left(g^{rn}g_{uu,r}\right)_{,q}-\left(g^{rn}\left(2rd_q+c_q\right)\right)_{,u}-4\left(g^{ns}E_{s[u}\right)_{,q]}\right]\bigg]m_i^pm_j^q
\end{eqnarray}
where $\tilde{C}_{pqnm}$ is the Weyl tensor equation for the obtained metric functions. We can conclude that, RT spacetime algebraic classification is not revealed analytically for this expansion scalar and these metric functions, as $D\rightarrow \infty$.

On the other hand, we can determine the classification of RT spacetime with the same expansion scalar $\Theta=\frac{1}{r}$ for off-diagonal terms ($g_{up}$) vanish. According to this simplification, $d_p=c_p=0$ and the metric coefficients become $g^{rp}=0$ and $g^{rr}=-g_{uu}$, thus, the metric (\ref{metric}) becomes,
\begin{equation}
	ds^2=r^2h_{pq}(u,x)dx^pdx^q-2dudr+g_{uu}(u,r,x)du^2.
\end{equation}
This spacetime is Type II or more special as the Weyl scalars $\Psi_{0^{ij}}=\Psi_{1T^{i}}=\Psi_{1^{ijk}}=0$ and it is algebraically special. Additionally, it will be Type II(a) or more special for the limit of $D\rightarrow \infty$ when the Weyl scalar $\Psi_{2S}=0$ which gives,
\begin{eqnarray}\label{eqn1}
	g_{uu}=-\frac{c_1(u,x)}{3r^3}+c_2(u,x).
\end{eqnarray}
The spacetime will be Type II(b) or more special as the Weyl scalar $\Psi_{2T^{ij}}$ vanishes that yields,
\begin{eqnarray}
	h_{pq,u}=h_{pq}\left(-g_{uu,r}+2g_{uu}\right)
\end{eqnarray}
and it will occur Type II(ab) if the metric function reads to,
\begin{eqnarray}\label{eqn2}
	h_{pq,u}=h_{pq}\left(-\frac{c_1(u,x)}{r^4}\left(1+\frac{2r}{3}\right)+2c_2(u,x)\right).
\end{eqnarray}
RT spacetime will be Type II(c) or more special as the Weyl scalar $\Psi_{2^{ijk\ell}}$ vanishes which reads;
\begin{eqnarray}
	~^sR_{pqmn}=r^3\left[h_{nq}\left(\frac{h_{pm,u}}{4}+g_{uu}h_{pm}\right)-h_{mq}\left(\frac{h_{pn,u}}{4}+g_{uu}h_{pn}\right)\right]
\end{eqnarray}
and it will become Type II(abc) if  
\begin{eqnarray}\label{eqn3}
	~^sR_{pqnm}=\frac{r^3}{4}\left(h_{nq}h_{pm}-h_{mq}h_{pn}\right)\left(\frac{c_1(u,x)}{r^4}\left(1+2r\right)+6c_2(u,x)\right).
\end{eqnarray}
When we set $\Psi_{2^{ij}}=0$, the spacetime will be Type II(d) or more special and we get $h_{qm,u}=h_{pm,u}$. Without loss of generality we can choose the metric coefficient $h_{pq}(u,x)\rightarrow h_{pq}(u)$ for the Type II(abcd) (or Type III) with these results. Further, this result satisfy RT spacetime is Type III(b) while the component of Weyl scalar $\Psi_{3^{ijk}}=0$. RT will be Type III(a) and N with the condition of  $\Psi_{3T^{i}}=0\rightarrow \left(rg_{uu,p}\right)_{,r}=0\rightarrow c_1(u,x)\rightarrow c_1(u), c_2(u,x)\rightarrow c_2(u)$.
It will be Type O when $\Psi_{4^{ij}}=0$ which gives,
\begin{eqnarray}
	h_{pq}\left(\frac{5}{6}c_1(u)-r^3c_2(u)\right)=r^{6}g^{\ell s}h_{ps}h_{\ell q}.
\end{eqnarray}

Also, we can introduce classification for the secondary WAND $\mathbb{\ell}$. RT spacetime will be Type $I_{i}$ and $II_{i}$ only the component of Weyl scalar $\Psi_{4^{ij}}$ vanishes and it yields;
\begin{eqnarray}\label{eqn4}
	\frac{rg_{uu}}{2}\left(h_{pq}g_{uu,r}+h_{pq,u}\right)+\frac{r^2}{2}h_{pq,u}g_{uu,r}-r^4g^{\ell s}h_{ps,u}h_{\ell q,u}-\frac{r^2h_{pn}}{2}\left(\left(-g^{ns}g_{uu,s}\right)_{,q}+r^2\left(g^{ns}h_{pq,u}\right)_{,u}\right)=0.~~~~~
\end{eqnarray}
If the metric functions satisfy equations (\ref{eqn1}, \ref{eqn2}, \ref{eqn3}) and $h_{pq}(u,x)\rightarrow h_{pq}(u)$ with equation (\ref{eqn4}) RT spacetime is Type $III_i$ for secondary WAND $\mathbf{\ell}$.

\section{Conclusion}\label{s5}
Algebraic classification of the higher dimensional RT and Kundt spacetimes were investigated with the method of taking the limit of dimension of spacetime $D\rightarrow \infty$. The results supported the method that the limit of dimension goes to infinity which simplify the theory of general relativity. In general, nor at this limit or any dimension $D>4$, without any restrictions, classification of  RT spacetime which is Type I(b) unchanged. Also the spacetime is not algebraically special and the primary WAND $\mathbf{k}$ is a WAND because all +1 components of boost weight do not vanish. 

As Kundt spacetime is corresponding $\Theta=0$ condition of RT spacetime, in general, at least it is Type I(b). We obtained other types and subtypes by setting the components of Weyl scalar are zero.When the all metric functions are chosen independent of parameter r, which is matching pp-waves, the spacetime became Type II(abd). Restrictions and the matching types and subtypes are summarized at the Table \ref{table2} for this case. 

Although, general classification of RT spacetime with necessary vanishing components of Weyl scalar for both WANDs was given at Table \ref{table2}, several different cases were discussed at the last chapter. When we set the $\Theta=\frac{1}{r}$ and off-diagonal terms $g_{up}=0$, the spacetime became algebraically special and it is Type II or more special. 

This paper was prepared to fill the gap in the literature by analyzing the algebraic classification of RT spacetime with the limit of dimension $D\rightarrow 0.$ In the future, the special types and subtypes can be studied by solving field equations which helps to understand the limitation method.

\appendix
\section{Curvature Tensors of General RT Spacetime}\label{appendix}
The non-zero Christoffel symbols of the metric (\ref{metric}) are obtained;
\begin{eqnarray}
	\Gamma^{u}_{~uu}&=&\frac{1}{2}g_{uu,r}\\
	\Gamma^{u}_{~up}&=&\frac{1}{2}g_{up,r}\\
	\Gamma^{u}_{~pq}&=&\Theta g_{pq}\\
	\Gamma^{r}_{~ur}&=&\frac{1}{2}\left(g^{rp}g_{up,r}-g_{uu,r}\right)\\
	\Gamma^{r}_{~up}&=&\frac{1}{2}\left(-g^{rr}g_{up,r}-g_{uu,p}+2g^{rn}E_{pn}\right)\\
	\Gamma^{r}_{~uu}&=&\frac{1}{2}\left(g^{rr}g_{uu,r}-g_{uu,u}+2g^{rn}E_{un}\right)\\
	\Gamma^{r}_{~pq}&=&-\Theta g^{rr}+\frac{1}{2}g_{pq,u}-g_{u(p,q)}+g_{un}~ ^s\Gamma^{n}_{~pq}\\
	\Gamma^{r}_{~rp}&=&\frac{1}{2} \left(2\Theta g_{um}-g_{up,r}\right)\\
	\Gamma^{p}_{~uu}&=&\frac{1}{2}\left(-g^{rp}g_{uu,r}-2g^{pn}E_{un}\right)\\
	\Gamma^{p}_{~ur}&=&\frac{1}{2}g^{pq}g_{uq,r}\\
	\Gamma^{p}_{~uq}&=&\frac{1}{2}\left(-g^{rp}g_{uq,r}+2E_{nq}\right)\\
	\Gamma^{p}_{~rq}&=&\Theta \delta^{p}_{~q}\\
	\Gamma^{m}_{~pq}&=&-\Theta g_{up}+~^s\Gamma^{m}_{~pq}
\end{eqnarray}
where $^s\Gamma^{n}_{~pq}$ is the chrisstoffel symbol of the spatial metric $g_{pq}$ and,
\begin{eqnarray}
	E_{pq}&=&g_{u[p,q]}+\frac{1}{2} g_{pq,u}\\
	E_{up}&=&g_{u[p,u]}+\frac{1}{2} g_{up,u}.
\end{eqnarray}
The Rieman tensors of the metric (\ref{metric}) are ;
\begin{eqnarray}
	R_{prrq}&=&g_{pq}\left(\Theta_{,r}+\Theta^2\right)\\
	R_{ruur}&=& \frac{1}{2}g_{uu,rr}-\frac{1}{4}g^{pq}g_{up,r}g_{uq,r}\\
	R_{ruup}&=&g_{u[u,p]r}+\frac{1}{2}\Theta\left(2E_{up}-g_{up}g_{uu,r}\right)-\frac{1}{2}g^{mn}g_{um,r}E_{np}\nonumber\\
	&&+\frac{1}{4}g^{rm}g_{um,r}g_{up,r}\\
	R_{rurp}&=&-\frac{1}{2}g_{up,rr}+\frac{1}{2}\Theta g_{up,r}\\
	R_{rupq}&=&g_{u[p,q]r}+\Theta \left(g_{u[p}g_{q]u,r}+E_{qm}-E_{pm}\right)\\
	R_{prmq}&=&2\Theta^2g_{p[m}g_{q]u}+\Theta g_{p[q}g_{m]u,r}+g_{p[q}\Theta_{,m]}\\
	R_{pruq}&=&-\frac{1}{2}g_{pn}g^{ms}g_{us,r} ~^s\Gamma^{n}_{~mq}-\frac{1}{4}g_{up,r}g_{uq,r}-\frac{1}{2}g_{pn}\left(g^{nm}g_{um,r}\right)_{,q}+g_{pq}\Theta_{,u}\nonumber\\
	&&+\frac{\Theta}{2}\left(-g_{pq}g^{rm}g_{um,r}+g_{pq}g_{uu,r}+ g_{uq}g_{up,r}+2E_{qp}\right)~~~~~~~~\\
	R_{pumq}&=&g_{up}g_{u[q,m]r}+g_{pn}g^{rs} ~^s\Gamma^{n}_{~s[q}g_{m]u,r}-g_{pn}g^{s\ell}E_{\ell[m} ~^s\Gamma^{n}_{~q]s}\nonumber\\
	&&-E_{p(q}g_{m)u,r}+g_{pn}\left(\left(g^{rn}g_{u[m,r}\right)_{,q]}+2\left(g^{n\ell}E_{\ell[m}\right)_{,q]}\right)\nonumber\\
	&&+\Theta\left(g_{uu,[m}g_{q]p}+g^{rr}g_{p[q}g_{m]u,r}+2g^{rs}E_{s[m}g_{q]p}\right)~\\
	R_{puqu}&=&g_{up}g_{u[u,q]r}-\frac{g_{pn}}{2}\left[\left(g^{rn}g_{uu,r}\right)_{,q}-\left(g^{rn}g_{uq,r}\right)_{,u}-2\left(g^{ns}E_{us}\right)_{,q}+2\left(g^{ns}E_{sq}\right)_{,u}\right]\nonumber\\
	&&+E_{p[q}g_{u]u,r}+\frac{1}{2}g^{rs}g_{uq,r}E_{ps}-g^{\ell s}E_{ps}E_{\ell q}+\frac{g_{pn}~^s\Gamma^{n}_{~sq}}{2}\left(g^{rs}g_{uu,r}+2g^{st}E_{ut}\right)\nonumber\\
	&&+\frac{1}{4}g_{up,r}g_{uu,q}+\frac{1}{4}g^{rr}g_{up,r}g_{uq,r}-\frac{1}{2}g^{r\ell}g_{up,r}E_{\ell q}\nonumber\\
	&&+\frac{\Theta}{2}\left(-2g_{up}E_{uq}+\frac{g_{pq}}{2}\left(-g_{uu,u}-g^{rr}g_{uu,r}+2g^{r\ell}E_{u\ell}\right)\right)~~~~~~\\
	R_{pqmn}&=&~^sR_{pqmn}+g_{ps}g^{rt}~^{s}\Gamma^s_{~t[n}g_{m]q,r}+2g_{pq}\left(\Theta g_{u[m}\right)_{,n]}+g_{up}g_{q[n,m]r}\nonumber\\
	&&+2\Theta g_{pq}g_{u[n,m]}+\Theta g_{up}g_{u[n}g_{m]q,r}+\Theta \left(g_{pn}\tilde{E}_{qm}-g_{pm}\tilde{E}_{qn}\right)\nonumber\\
	&&+\Theta g^{rr}g_{p[n}g_{m]q,r}+2g_{us}q_{p[m}~^s\Gamma^s_{~n]q}+\Theta g_{pq}g_{u[m}g_{n]u,r}
\end{eqnarray}
where $\tilde{E}_{pq}=g_{u(p,q)}-\frac{1}{2}g_{pq,u}$.

Ricci tensors become;
\begin{eqnarray}
	R_{rr}&=&-\left(D-2\right)\left(\Theta^2+\Theta_{,r}\right)\\
	R_{rp}&=& -\frac{1}{2}g_{up,rr}+\Theta_{,r}g_{up}+\left(D-2\right)\Theta^2g_{up}-\frac{D-3}{2}\Theta_{,p}-\frac{D-4}{2}\Theta g_{up,r}~~~~~~~~~~\\
	R_{ru}&=&-\frac{1}{2}g_{uu,rr}+\frac{1}{2}g^{rp}g_{up,rr}-g^{ms}g_{us,r}~^s\Gamma^{q}_{~mq}-\frac{1}{2}\left(g^{mq}g_{um,r}\right)_{,q}+g^{pq}\Theta E_{qp}\nonumber\\
	&&+\frac{D-2}{2}\left(-\Theta g^{rp}g_{up,r}+\Theta g_{uu,r}+2 \Theta_{,u}\right)\\
	R_{uu}&=&-\frac{1}{2}g^{rr}g_{uu,rr}-g^{rp}g_{u[u,p]r}+\frac{1}{2}g^{pq}g_{up,r}g_{uq,r}-\frac{1}{2}g^{rp}g^{rs}g_{up,r}g_{us,r}\nonumber\\
	&&+\frac{1}{2}g^{pq}g^{rs}g_{up,r}E_{sq}-\frac{1}{2}\left(\left(g^{rq}g_{uu,r}\right)_{,q}-\left(g^{rq}g_{uq,r}\right)_{,u}-2\left(g^{qs}E_{us}\right)_{q}+2\left(g^{qs}E_{sq}\right)_{,u}\right)\nonumber\\
	&&+g^{pq}E_{p[q}g_{u]u,r}+\frac{1}{2}g^{pq}g^{rs}g_{uq,r}E_{ps}-g^{pq}g^{\ell s}E_{ps}E_{\ell q}+\frac{~^s\Gamma^{q}_{~sq}}{2}\left(g^{rs}g_{uu,r}+2g^{s\ell} E_{u\ell}\right)\nonumber\\
	&&+\frac{1}{4}g^{pq}g_{up,r}g_{uu,q}+\frac{\Theta}{2}\left(-\left(D-2\right)g_{uu,u}-\left(D-2\right)g^{rr}g_{uu,r}\right)+\left(D-4\right)g^{rs}E_{us}~~~~~~~~~\\
	R_{up}&=&-2g_{u[u,p]r}-\frac{1}{2}g^{rr}g_{up,rr}+g^{rq}g_{u[q,p]r}-\frac{1}{2}g^{rq}g_{up,r}g_{uq,r}+g_{up}\Theta_{,u}+\left(g^{rp}g_{u[q,r}\right)_{,p]}\nonumber\\
	&&-\frac{1}{2}g^{rq}g_{pn}\left(g^{nm}g_{um,r}\right)_{,q}-\Theta\left(E_{up}-g^{rp}\left(E_{pm}-E_{qm}\right)\right)-\frac{1}{2}g^{mq}E_{mq}g_{up,r}\nonumber\\
	&&+2 g^{qm}\left(g^{n\ell} E_{\ell[q}\right)_{,p]}-g^{s\ell}E_{\ell [q}~^s\Gamma^q_{~ps]}-\frac{1}{2}g^{rq}g_{pn}g^{ms}g_{us,r}~^{s}\Gamma^n_{~mq}+g^{uq}g_{up,rs} ~^s\Gamma^q_{~s[p}g_{q]u,r}\nonumber\\
	&&+\Theta g^{rp}\left(g_{uq}g_{up,r}-\frac{1}{2}g_{up}g_{uq,r}\right)\nonumber\\
	&&-\frac{\Theta}{2}\left(\left(D-3\right)g_{uu,p}+\left(D-4\right)g^{rr}g_{up,r}+2\left(D-4\right)g^{rs}E_{sp}\right)\\
	R_{pq}&=&~^sR_{pq}-g_{pq}\left(-2\Theta_{,u}+g^{rn}\Theta_{,n}+g^{rr}\Theta_{,r}\right)+\Theta^2\left(g_{pq}\left(g_{uu}+g^{rn}g_{un}\right)-2g_{up}g_{uq}\right)\nonumber\\
	&&+\Theta\left[g_{pq}\left(-2g^{rn}g_{un,r}-g_{uu,r}\right)+g_{up}g_{uq,r}+2g_{u[q,p]}+\frac{1}{2}g^{rn}g_{uq}g_{np,r}\right]\nonumber\\
	&&-\frac{1}{2}g_{up,r}g_{uq,r}-g^{rn}g_{n(p}\Theta_{,q)}-g^{ms}g_{us,r}~^s\Gamma^n_{~m(q}g_{p)n}+g^{rn}g_{p[q,n]r}-g_{pq,r}g_{uu}\nonumber\\
	&&-\frac{D-2}{2}g^{rr}g_{pq,r}-(D-3)\Theta \tilde{E}_{pq}+(D-3)g_{us}~^s\Gamma^s_{~pq}
\end{eqnarray}

Ricci scalar becomes;

\begin{eqnarray}
	R&=&~^{s}R+g_{uu,rr}-2g_{un,rr}+g^{pq} g^{rn}g_{p[q,n]r}+g^{pn}g_{un,r}~^{s}\Gamma^{q}_{~mq}+\left(g^{mq}g_{um,r}\right)_{,q}+2g^{rp}g_{up}\Theta_{,r}\nonumber\\
	&&-\frac{1}{2}g^{pq}g_{up,r}g_{uq,r}-2\Theta g_{uu}+g^{pq}\Theta\left(-2E_{qp}+g_{up}g_{uq,r}+2g_{u[q,p]}+\frac{1}{2}g^{rn}g_{uq}g_{up,r}\right)\nonumber\\
	&&-\Theta\left(2\left(D-3\right)g^{rn}g_{un,r}+2\left(D-2\right)g_{uu,r}+\left(D-2\right)g^{rr}+\left(D-3\right)\tilde{E}_{pq}g^{pq}\right)\nonumber\\
	&&-2\left(D-2\right)g^{rr}\Theta_{,r}-2\left(D-2\right)g^{rn}\Theta_{,n}++\Theta^2\left(2\left(D-2\right)g_{uu}+2\left(D-3\right)g^{rn}g_{un}\right)\nonumber\\
	&&+\left(D-3\right)g^{pq}g_{us}~^{s}\Gamma^{s}_{~pq}
\end{eqnarray}

As dimension $D\rightarrow\infty$ the Weyl tensor of the metric (\ref{metric}) becomes;
\begin{eqnarray}
	C_{rprq}&=&0,\\
	C_{rpru}&=&-\frac{1}{2}g_{up,rr}+\Theta g_{up,r}+\frac{\Theta_{,p}}{2}+g_{up}\Theta_{,r}\\
	C_{prmq}&=&0,\\
	C_{ruru}&=&-\frac{1}{4}g_{uu,rr}+\frac{1}{4}g^{pq}g_{up,r}g_{uq,r}+g_{uu}\Theta^2+g_{uu}\Theta_{,r}-\Theta g_{uu,r}+\Theta g^{rp}g_{up,r}-2 \Theta_{,u}~~~~~~~~~~\\
	C_{rpuq}&=&\frac{1}{2}g_{pn}g^{ms}g_{us,r}~^{s}\Gamma^{n}_{~mq}+\frac{1}{4}g_{up,r}g_{uq,r}+\frac{1}{2}g_{pn}\left(g^{nm}g_{um,r}\right)_{,q}-g_{pq}\Theta_{,u}-\frac{g_{up}}{2}\Theta_{,q}\nonumber\\
	&&-\frac{\Theta}{2}\left(-g_{pq}g^{rm}g_{um,r}+g_{pq}g_{uu,r}+2g_{u(p}g_{q)u,r}+2E_{qp}\right)+\Theta^2g_{up}g_{uq}-\frac{g^{rr}}{2}g_{pq,r}\\
	C_{rupq}&=&g_{u[p,q]r}-g_{u[p}\Theta_{,q]}+\Theta\left(E_{qm}-E_{pm}\right)\\
	C_{pqmn}&=&~^sR_{pqmn}+g_{ps}g^{rt}~^s\Gamma^s_{~t[n}g_{m]q,r}+g_{us}~^s\Gamma^{s}_{~p[n}g_{m]q}+2 g_{pq}\left(\Theta g_{u[m}\right)_{,n]}+g_{up}g_{q[n,m]r}\label{cpqmn}\nonumber\\
	&&+\Theta\left( g_{pq}\left(2g_{u[n,m]}+g_{u[m}g_{n]u,r}\right)+g_{up}g_{u[n}g_{m]q,r}-\tilde{E}_{p[n}g_{m]q}\right)-g^{rr}g_{q[m}g_{n]p,r}\\
	C_{ruup}&=&g_{u[u,p]r}-\frac{\Theta}{2}\left(\left(g_{uu}g_{up}\right)_{,r}+g_{uu,p}+g^{rr}g_{up,r}+2g^{rs}E_{sp}\right)+g_{uu}\left(\Theta^2g_{up}-\Theta_{,p}\right)\nonumber\\
	&&-\frac{1}{2}g^{mn}g_{um,r}E_{np}+\frac{1}{4}g_{um,r}g_{up,r}\\
	C_{upmq}&=&-g_{up}g_{u[q,m]r}-g_{pn}g^{rs}~^s\Gamma^n_{~s[q}g_{m]u,r}+g_{pn}g^{s\ell}E_{\ell[m}~^s\Gamma^n_{~q]s}+E_{p(q}g_{m)u,r}\nonumber\\
	&&-g_{pn}\left(\left(g^{rn}g_{u[m,r}\right)_{,q}+2\left(g^{n\ell}E_{\ell[m}\right)_{,q]}\right)+g^{rr}g_{u[m}g_{q]p,r}\\
	C_{upuq}&=&g_{up}g_{u[u,q]r}-\frac{g_{pn}}{2}\left[\left(g^{rn}g_{uu,r}\right)_{,q}-\left(g^{rn}g_{uq,r}\right)_{,u}-2\left(g^{ns}E_{us}\right)_{,q}+2\left(g^{ns}E_{sq}\right)_{,u}\right]\nonumber\\
	&&+E_{p[q}g_{u]u,r}+\frac{1}{2}g^{rs}g_{uq,r}E_{ps}-g^{\ell s}E_{ps}E_{\ell q}+\frac{g_{pn}~^s\Gamma^{n}_{~sq}}{2}\left(g^{rs}g_{uu,r}+2g^{st}E_{ut}\right)\nonumber\\
	&&+\frac{1}{4}g_{up,r}g_{uu,q}+\frac{1}{4}g^{rr}g_{up,r}g_{uq,r}-\frac{1}{2}g^{r\ell}g_{up,r}E_{\ell q}+\frac{1}{2}g^{rr}g_{uu}g_{pq,r}\nonumber\\
	&&+\Theta\left(-g_{up}E_{uq}-g_{uu,(p}g_{q)u}-g^{rr}g_{u(p}g_{q)u,r}+2g^{rs}E_{s(p}g_{q)u}\right)
\end{eqnarray}

\acknowledgments

PK was supported by postdoctoral grant of Scientific and Technological Research
Council of Turkey (TÜBİTAK) project with ID:1059B191801670. PKU is indebted to Roberto Emparan to happen this project.



\begin{thebibliography}{99}

\bibitem{Emparan:2013moa} R.~Emparan, R.~Suzuki and K.~Tanabe,
\emph{The large D limit of General Relativity,
JHEP }.\textbf{06} (2013), 009.

\bibitem{Emparan:2013xia}
R.~Emparan, D.~Grumiller and K.~Tanabe,
\emph{Large-D gravity and low-D strings,
Phys. Rev. Lett.} \textbf{110} (2013) no.25, 251102.

\bibitem{Emparan:2014cia}
R.~Emparan and K.~Tanabe,
\emph{Universal quasinormal modes of large D black holes,
Phys. Rev. D} \textbf{89} (2014) no.6, 064028.

\bibitem{Emparan:2015rva}
R.~Emparan, R.~Suzuki and K.~Tanabe,
\emph{Quasinormal modes of (Anti-)de Sitter black holes in the 1/D expansion, JHEP} \textbf{04} (2015), 085.

\bibitem{Emparan:2015gva}
R.~Emparan, R.~Suzuki and K.~Tanabe,
\emph{Evolution and End Point of the Black String Instability: Large D Solution, Phys. Rev. Lett.} \textbf{115} (2015) no.9, 091102.

\bibitem{Andrade:2019edf}
T.~Andrade, R.~Emparan, D.~Licht and R.~Luna,
\emph{Black hole collisions, instabilities, and cosmic censorship violation at large $D$,	JHEP }\textbf{09} (2019), 099.

\bibitem{Licht:2020odx}
D.~Licht, R.~Luna and R.~Suzuki,
\emph{Black Ripples, Flowers and Dumbbells at large $D$,
JHEP} \textbf{04}, (2020), 108. 

\bibitem{Suzuki:2020kpx}
R.~Suzuki,
\emph{Black hole interactions at large $D$: brane blobology,
JHEP} \textbf{02} (2021), 131. 

\bibitem{Suzuki:2021lrw}
R.~Suzuki and S.~Tomizawa,
\emph{Squashed black holes at large D,
JHEP} \textbf{12} (2021), 194. 

\bibitem{Licht:2022wmz}
D.~Licht, R.~Luna and R.~Suzuki,
\emph{Lattice Black Branes at Large $D$,}
[arXiv:2201.11687 [hep-th]].

\bibitem{Colin-Ellerin:2019vst}
S.~Colin-Ellerin, V.~E.~Hubeny, B.~E.~Niehoff and J.~Sorce,
\emph{Large-$d$ phase transitions in holographic mutual information, JHEP} \textbf{20} (2020), 173.
\bibitem{RT60} Robinson I. and Trautman A. 1960. \emph{Spherical gravitational waves Phys. Rev. Lett.} \textbf{4} 431–2.
\bibitem{RT62}Robinson I. and Trautman A. 1962. \emph{Some spherical gravitational waves in general relativity Proc.	Roy. Soc. A .}\textbf{265} 463–73.

\bibitem{Kundt61} Kundt W. .\emph{ The plane-fronted gravitational waves. Z. Physik,} \textbf{163} (1961) 77–86.
\bibitem{Kundt62} Kundt W. \emph{Exact solutions of the field equations: twist-free pure radiation fields. Proc. Roy.Soc. A.} \textbf{270}.  ()1962). 328–34. 


\bibitem{Podolsky:2006du}
J.~Podolsky and M.~Ortaggio,
\emph{Robinson-Trautman spacetimes in higher dimensions,
Class. Quant. Grav.} \textbf{23},  (2006) 5785-5797.

\bibitem{Ortaggio:2007hs}
M.~Ortaggio, J.~Podolsky and M.~Zofka,
\emph{Robinson-Trautman spacetimes with an electromagnetic field in higher dimensions, Class. Quant. Grav.} \textbf{25} (2008), 025006.

\bibitem{Ortaggio:2014gma}
M.~Ortaggio, J.~Podolsk\'y and M.~\v{Z}ofka,
\emph{Static and radiating p-form black holes in the higher dimensional Robinson-Trautman class, JHEP }\textbf{02}, 045 (2015) doi:10.1007/JHEP02(2015)045 [arXiv:1411.1943 [gr-qc]].

\bibitem{Podolsky:2008ec}
J.~Podolsky and M.~Zofka,
\emph{General Kundt spacetimes in higher dimensions,
Class. Quant. Grav.} \textbf{26}, 105008 (2009)
doi:10.1088/0264-9381/26/10/105008 [arXiv:0812.4928 [gr-qc]].

\bibitem{Petrov} Petrov, A.Z. . \emph{Classification of spaces defined by gravitational fields. Uch. Zapiski.Kazan Gos. Univ. }144, (1954) 55.

\bibitem{Milson:2004jx}
R.~Milson, A.~Coley, V.~Pravda and A.~Pravdova,
\emph{Alignment and algebraically special tensors in Lorentzian geometry,
Int. J. Geom. Meth. Mod. Phys. }\textbf{2}, 41-61 (2005)
doi:10.1142/S0219887805000491
[arXiv:gr-qc/0401010 [gr-qc]].

\bibitem{Podolsky:2014mpa} 
J.~Podolsky and R.~Svarc,
\emph{Algebraic structure of Robinson-Trautman and Kundt geometries in arbitrary dimension, Class.\ Quant.\ Grav.\ } {\bf 32}, 015001 (2015)
doi:10.1088/0264-9381/32/1/015001
[arXiv:1406.3232 [gr-qc]]. 	R.~\v{s}varc and J.~Podolsk\'y,
\emph{``Algebraic aspects of general non-twisting and shear-free spacetimes,''} doi:10.1142/9789813226609\_0301

\bibitem{Podolsky:2013qwa}
J.~Podolsky and R.~Švarc,
\emph{Explicit algebraic classification of Kundt geometries in any dimension,
Class. Quant. Grav.} \textbf{30} (2013), 125007.

\bibitem{Stephani} Stephani H, Kramer D, MacCallum M, Hoenselaers C. and Herlt E. \emph{ Exact Solutions of
Einstein’s Field Equations.} Cambridge: Cambridge University Press (2003).

\bibitem{Griffits} Griffiths J. B. and Podolsky J. \emph{ Exact Space-Times in Einstein’s General Relativity}, Cambridge: Cambridge University Press (2009).

\bibitem{Durkee:2011tx}
M. Durkee,
\emph{New approaches to higher-dimensional general relativity,}
doi:10.17863/CAM.16104
[arXiv:1104.4414 [gr-qc]].



\bibitem{classification} Coley A, Milson R, Pravda V. and Pravdov A. \emph{Classification of the Weyl tensor in higher
dimensions. Class. Quantum Grav.} \textbf{21} (2004) L35–41.

\bibitem{Brinkmann:1925fr}
H.~W.~Brinkmann,
\emph{Einstein spapces which are mapped conformally on each other,
Math. Ann. }\textbf{94}, (1925), 119-145 
doi:10.1007/BF01208647




\end{thebibliography}
\end{document}